\documentclass{llncs}
\usepackage{mathptmx}       
\usepackage{helvet}         
\usepackage{courier}        
\usepackage{type1cm}        
\usepackage{makeidx}         
\usepackage{graphicx}        
\usepackage{multicol}        
\usepackage[bottom]{footmisc}
\usepackage{subfigure}
\usepackage{amsfonts}
\usepackage[cmex10]{amsmath}
\usepackage{hyperref}
\usepackage{multicol}
\usepackage{cite}
\usepackage{sidecap}

\begin{document}

\title{Detecting Global Community Structure in a COVID-19 Activity Correlation Network}
\titlerunning{Detecting Global Community Structure in COVID-19 Activities}  
%
\author{Hiroki Sayama\inst{1,2}}
\authorrunning{Hiroki Sayama} 
%
\tocauthor{Hiroki Sayama}
\institute{Binghamton University, State University of New York, Binghamton, NY 13902-6000, USA,\\
	\email{sayama@binghamton.edu},\\ WWW home page:
	\texttt{\url{https://bingweb.binghamton.edu/~sayama/}}
	\and
	Waseda University, Shinjuku, Tokyo 169-8050, Japan}

\maketitle    

\begin{abstract}
The global pandemic of COVID-19 over the last 2.5 years have produced
an enormous amount of epidemic/public health datasets, which may also
be useful for studying the underlying structure of our globally
connected world. Here we used the Johns Hopkins
University COVID-19 dataset to construct a correlation network of
countries/regions and studied its global community
structure. Specifically, we selected countries/regions that had at
least 100,000 cumulative positive cases from the dataset and generated
a 7-day moving average time series of new positive cases reported for
each country/region. We then calculated a time series of daily change
exponents by taking the day-to-day difference in log of the number of
new positive cases. We constructed a correlation network by connecting
countries/regions that had positive correlations in their daily change
exponent time series using their Pearson correlation coefficient as
the edge weight. Applying the modularity maximization method revealed
that there were three major communities: (1) Mainly Europe + North
America + Southeast Asia that showed similar six-peak patterns during
the pandemic, (2) mainly Near/Middle East + Central/South Asia +
Central/South America that loosely followed Community 1 but had a notable increase of activities because of the Delta variant and was later impacted significantly by the Omicron variant, and (3)
mainly Africa + Central/East Canada + Australia that did not have much
activities until a huge spike was caused by the Omicron variant. These
three communities were robustly detected under varied
settings. Constructing a 3D ``phase space'' by using the median curves
in those three communities for $x$-$y$-$z$ coordinates generated an
effective summary trajectory of how the global pandemic progressed.
\end{abstract}

\section{Introduction}

The global pandemic of COVID-19 over the last 2.5 years have produced an enormous amount of epidemic/public health datasets, including the numbers of positive cases, deaths, hospitalizations and vaccine administrations \cite{JHU-dashboard,JHUgithub,CDCdata,WHO,NYtimesdata,worldometer}, individual-level case data \cite{xu2020,global-health}, and other forms of publicly available datasets \cite{cheng2020,shuja2021}. These datasets have been utilized extensively for modeling and visualization of the COVID-19 pandemic \cite{JHU-dashboard,CDCdata,WHO,NYtimesdata,coronatracker-paper,coronatracker,covidvisualizer} (including some of the author's own work as well \cite{sayama2020,sayama2021}), which have contributed significantly to both advancing scientific understanding of the disease spreading dynamics and also raising public awareness about the pandemic.

Although traditional pandemic data modeling, analysis and visualization typically consider the dynamics of epidemic variables in a population or a region in isolation, complex systems and network sciences have firmly established the need for taking the underlying network structure into account when considering epidemic dynamics \cite{pastor-satorras2001,moreno2002,pastor-satorras2015,masuda2017,thurner2020}. The majority of network epidemic research mostly focuses on the networks of social contacts and mobility connections among human individuals, which would have only limited connection to larger-scale epidemic/public health data described above. Meanwhile, network-oriented data modeling and analysis can be useful for understanding and prediction of disease spreading at such larger spatial scales. Examples include epidemic studies using global air traffic and tourism networks \cite{brockmann2013,tsiotas2022} and regional networks within a country \cite{dellarossa2020,amico2021}. 

Very little research has been done on network modeling and analysis of the COVID-19 data at a global scale. Earlier attempts include Zhu et al.'s work \cite{zhu2021} that conducted basic network modeling and analysis using correlations among countries in the first several months of the COVID-19 activities, although their network analysis remained rather elementary using only a simple thresholding method (i.e., correlations were not utilized quantitatively as edge weights) with no robustness check. Now that we have more than two years of detailed global pandemic data, this direction of research can be useful for detecting the underlying structure of our globally connected world in a more robust, quantitative manner. Over the last few years, we witnessed different parts of the world underwent different epidemic patterns, while some geographically distant regions showed similar activity patterns. Such differences and similarities of COVID-19 activities may help reveal the effective structure of international connectivities. In this paper, we constructed a correlation network of
countries/regions for the entire globe and studied its global community structure. We further used the detected major communities as a means to generate summary variables of the global epidemic state of the world, which allowed for low-dimensional ``phase space'' visualization of the COVID-19 epidemic dynamics.

\section{Dataset}

We used the COVID-19 time series dataset maintained and publicly released by the Center for Systems Science and Engineering (CSSE) at Johns Hopkins University \cite{JHU-dashboard,JHUgithub}. We obtained the time series data of the cumulative numbers of COVID-19 positive cases (written as $n^i_t$, where $i$ is the country/region and $t$ the date) for 285 countries/regions over the time period from January 22, 2020 to May 29, 2022. We selected only such countries/regions that had at least 100,000 cumulative positive cases as of May 29, 2022, in order to prevent the large stochasticity in smaller countries/regions from affecting our analysis. As a result, we had time series data for 139 countries/regions (listed in Table \ref{countries}), which was used for the network construction and analysis in this study. 

Note that the original data contained many reporting anomalies and other kinds of erratic values. To reduce the effects of such data anomalies on the results, the data was first smoothed over time and clipped to a finite value range before network modeling and analysis, as described in detail in the next section.

\begin{table}[ht]
\caption{List of 139 countries/regions that had at least 100,000 cumulative positive cases as of May 29, 2022. These countries/regions were used for network construction in this study. Note that some countries such as Australia and Canada are divided into multiple regions in the original Johns Hopkins University CSSE dataset, which we kept separated without aggregation in this study.}
{\scriptsize
\begin{multicols}{4}
\begin{description}
\item Afghanistan
\item Albania
\item Algeria
\item Argentina
\item Armenia
\item Australia: Australian Capital Territory
\item Australia: New South Wales
\item Australia: Queensland
\item Australia: South Australia
\item Australia: Tasmania
\item Australia: Victoria
\item Australia: Western Australia
\item Austria
\item Azerbaijan
\item Bahrain
\item Bangladesh
\item Belarus
\item Belgium
\item Bolivia
\item Bosnia and Herzegovina
\item Botswana
\item Brazil
\item Brunei
\item Bulgaria
\item Burma
\item Cambodia
\item Cameroon
\item Canada: Alberta
\item Canada: British Columbia
\item Canada: Manitoba
\item Canada: Ontario
\item Canada: Quebec
\item Canada: Saskatchewan
\item Chile
\item China: Hong Kong
\item China: Unknown
\item Colombia
\item Costa Rica
\item Croatia
\item Cuba
\item Cyprus
\item Czechia
\item Denmark
\item Dominican Republic
\item Ecuador
\item Egypt
\item El Salvador
\item Estonia
\item Ethiopia
\item Finland
\item France
\item France: Guadeloupe
\item France: Martinique
\item France: Reunion
\item Georgia
\item Germany
\item Ghana
\item Greece
\item Guatemala
\item Honduras
\item Hungary
\item Iceland
\item India
\item Indonesia
\item Iran
\item Iraq
\item Ireland
\item Israel
\item Italy
\item Jamaica
\item Japan
\item Jordan
\item Kazakhstan
\item Kenya
\item Kosovo
\item Kuwait
\item Kyrgyzstan
\item Laos
\item Latvia
\item Lebanon
\item Libya
\item Lithuania
\item Luxembourg
\item Malaysia
\item Maldives
\item Mauritius
\item Mexico
\item Moldova
\item Mongolia
\item Montenegro
\item Morocco
\item Mozambique
\item Namibia
\item Nepal
\item Netherlands
\item New Zealand
\item Nigeria
\item North Macedonia
\item Norway
\item Oman
\item Pakistan
\item Panama
\item Paraguay
\item Peru
\item Philippines
\item Poland
\item Portugal
\item Qatar
\item Romania
\item Russia
\item Rwanda
\item Saudi Arabia
\item Serbia
\item Singapore
\item Slovakia
\item Slovenia
\item South Africa
\item South Korea
\item Spain
\item Sri Lanka
\item Sweden
\item Switzerland
\item Taiwan
\item Thailand
\item Trinidad and Tobago
\item Tunisia
\item Turkey
\item Uganda
\item Ukraine
\item United Arab Emirates
\item United Kingdom
\item Uruguay
\item US
\item Uzbekistan
\item Venezuela
\item Vietnam
\item West Bank and Gaza
\item Zambia
\item Zimbabwe
\end{description}
\end{multicols}}
\label{countries}
\end{table}

\section{Methods}

To construct a correlation network of countries/regions with regard to their COVID-19 activity patterns, we first converted the time series for each country/region into another time series about the numbers of daily new positive cases and then smoothed it out with 7-day moving averaging to reduce day-to-day fluctuations and weekly cyclical effects. We then calculated yet another time series of {\em daily change exponents} by taking the day-to-day difference of $\log($number of new positive cases, 7-day averaged$)$. These can be summarized mathematically as follows:
\begin{eqnarray}
\text{Cumulative positive cases:} \quad & n^i_t & \\
\text{Daily new positive cases:} \quad & d^i_t & = n^i_t - n^i_{t-1}\\
\text{7-day average:} \quad & a^i_t & 
= \frac{1}{7} \sum_{j = 0}^6 d^i_{t-j}
\end{eqnarray}
\begin{eqnarray}
\text{Daily change exponent:} \quad & c^i_t & = \log(a^i_t) - \log(a^i_{t-1}) 
= \log \left( \frac{a^i_t}{a^i_{t-1}} \right)
\end{eqnarray}
We took this change exponent approach so as to avoid the results of correlation analysis being affected by the difference in the absolute volume of positive cases that varied greatly over space and time. The daily change exponent values were clipped to the $[-\alpha, \alpha]$ range with $\alpha = 7$ (this corresponds to $[e^{-7}, e^7] \sim [10^{-3}, 10^3]$ in linear daily change ratios) to prevent data reporting anomalies from influencing the correlation analysis too much.

\setcounter{footnote}{0}

We constructed a correlation network of countries/regions by measuring the Pearson correlation coefficients between their daily change exponent sequences (for the entire time period) and connecting such countries/regions that had a correlation coefficient greater than threshold $\rho$ (we used $\rho = 0$ in the main results), using the correlation coefficient as the edge weight. Any correlations with coefficients equal to or less than $\rho$ were ignored and not included as edges in the network\footnote{In the main result presented in this paper, the following countries/regions did not have any positive correlation with others so they do not show up in the correlation network and subsequent plots: Cambodia, Egypt, France: Martinique, Georgia, Kyrgyzstan, Luxembourg, and Oman.}. This made the countries/regions that exhibited similar COVID-19 activity patterns connected with each other with edge weights proportional to their correlation-based similarities. To study the global community structure of this network, we applied the modularity maximization method \cite{louvain,findgraphcommunities}\footnote{The modularity maximization method was chosen for community detection because we wanted to detect clusters of countries/regions that showed higher levels of correlation within themselves than across them. Meanwhile, other community detection methods can certainly be considered as well.}.All the network analyses and visualizations were conducted using Wolfram Research Mathematica 13.0.0.

\section{Results}

\begin{figure}[p]
\centering
\includegraphics[height=1.05\columnwidth,angle=90]{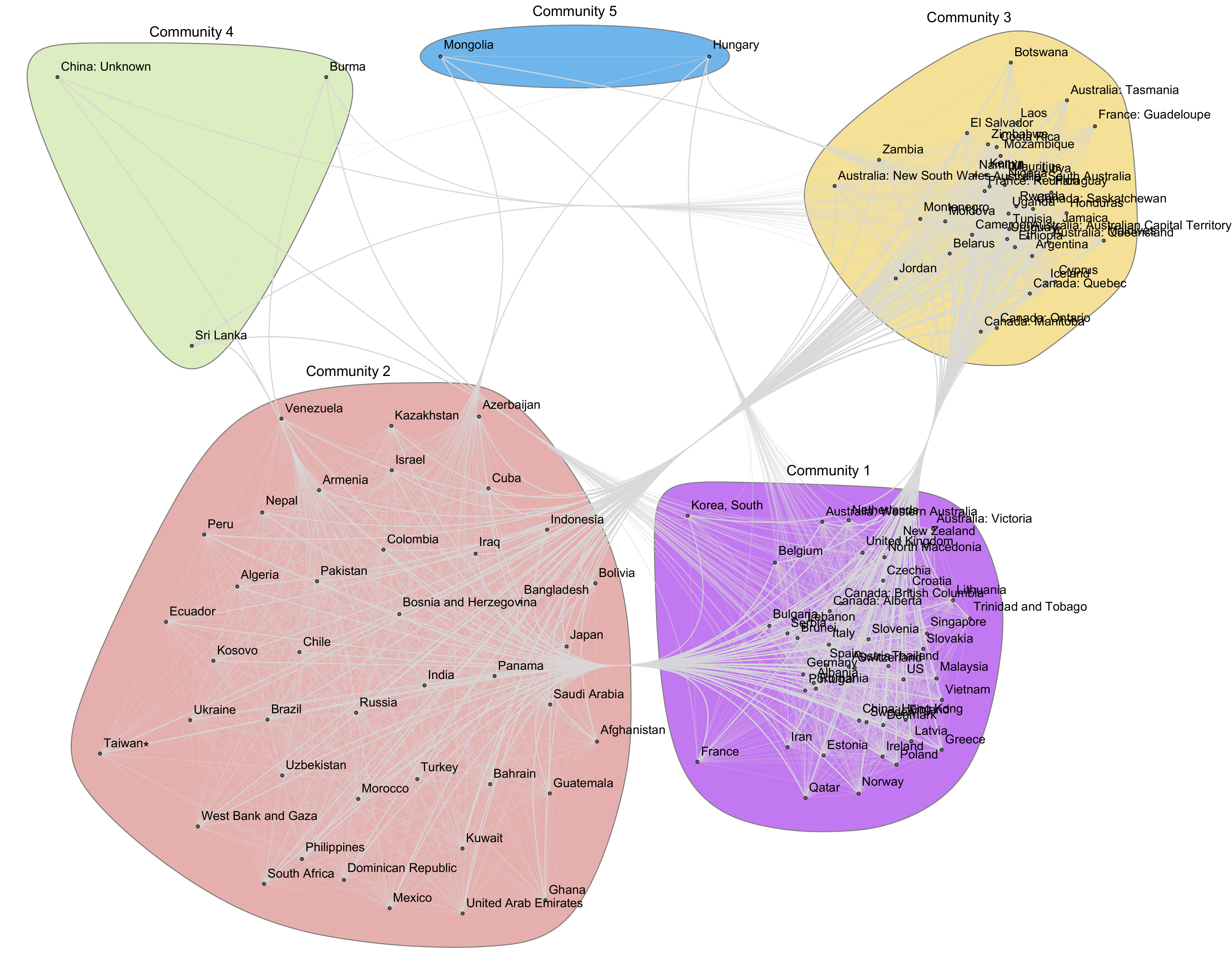}
\caption{(Rotated) Global community structure detected in the COVID-19 activity correlation network using the modularity maximization method \cite{louvain,findgraphcommunities}. Nodes represent countries/regions that had at least 100,000 cumulative positive cases as of May 29, 2022 (Table \ref{countries}) in the Johns Hopkins University COVID-19 data repository \cite{JHU-dashboard,JHUgithub}. Edges represent positive correlations of the COVID-19 activities in terms of the daily change exponent time series. Three major communities (1, 2, 3) were detected, together with two tiny ones (4, 5) in this example.}
\label{network-communities}
\end{figure}

\begin{figure}[p]
\centering
\begin{tabular}{lll}
\begin{minipage}[t]{0.55\columnwidth}
\vspace{0pt}
\includegraphics[width=\columnwidth]{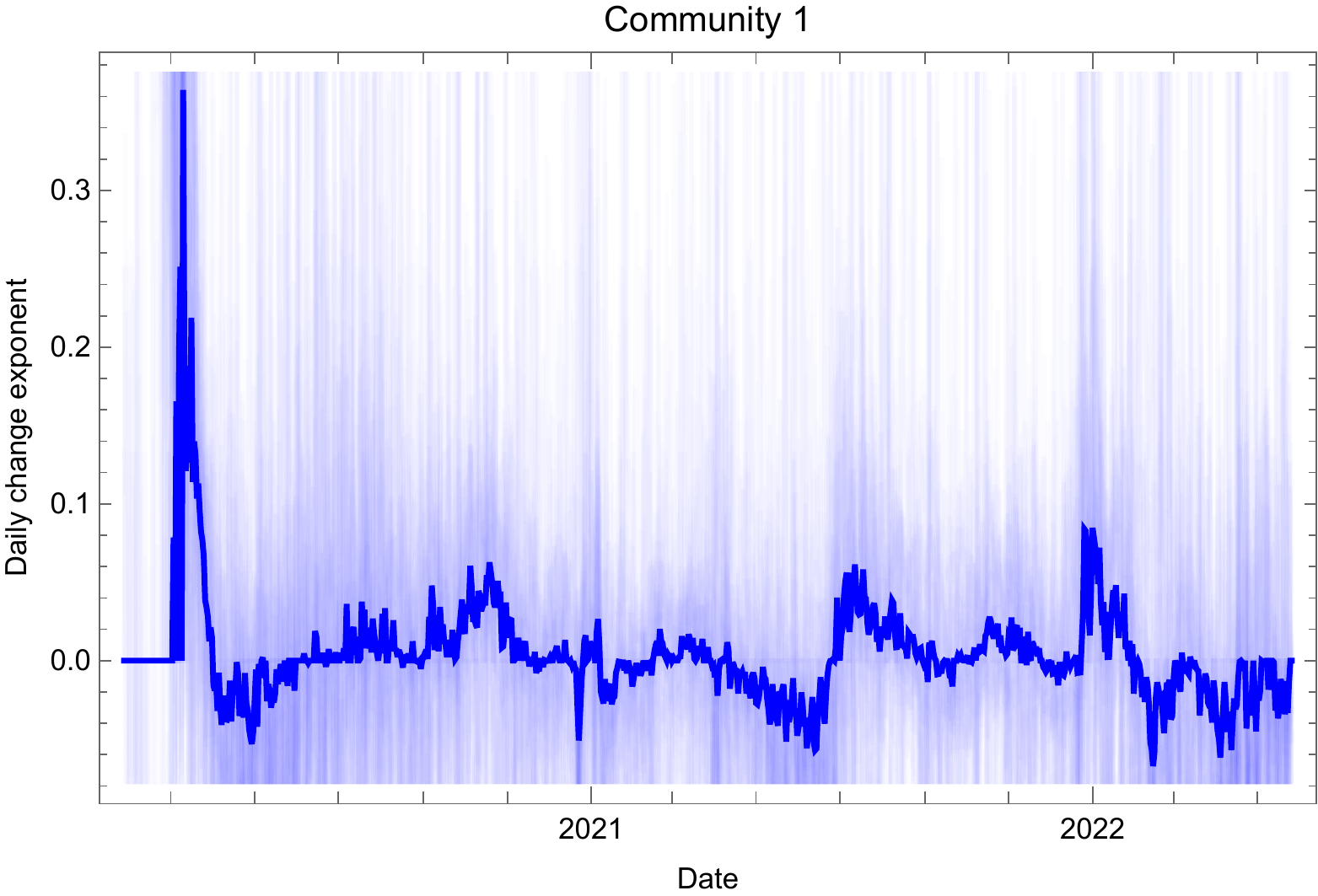} 
\vspace{0pt}
\end{minipage}
& ~ &
\begin{minipage}[t]{0.43\columnwidth}
\vspace{4pt}
{\scriptsize {\bf Community 1:} Albania, Australia: Victoria, Australia: Western Australia, Austria, Belgium, Brunei, Bulgaria, Canada: Alberta, Canada: British Columbia, China: Hong Kong, Croatia, Czechia, Denmark, Estonia, Finland, France, Germany, Greece, Iran, Ireland, Italy, Latvia, Lebanon, Lithuania, Malaysia, Netherlands, New Zealand, North Macedonia, Norway, Poland, Portugal, Qatar, Romania, Serbia, Singapore, Slovakia, Slovenia, South Korea, Spain, Sweden, Switzerland, Thailand, Trinidad and Tobago, United Kingdom, US, Vietnam}
\vspace{0pt}
\end{minipage}
\\
\begin{minipage}[t]{0.55\columnwidth}
\vspace{0pt}
\includegraphics[width=\columnwidth]{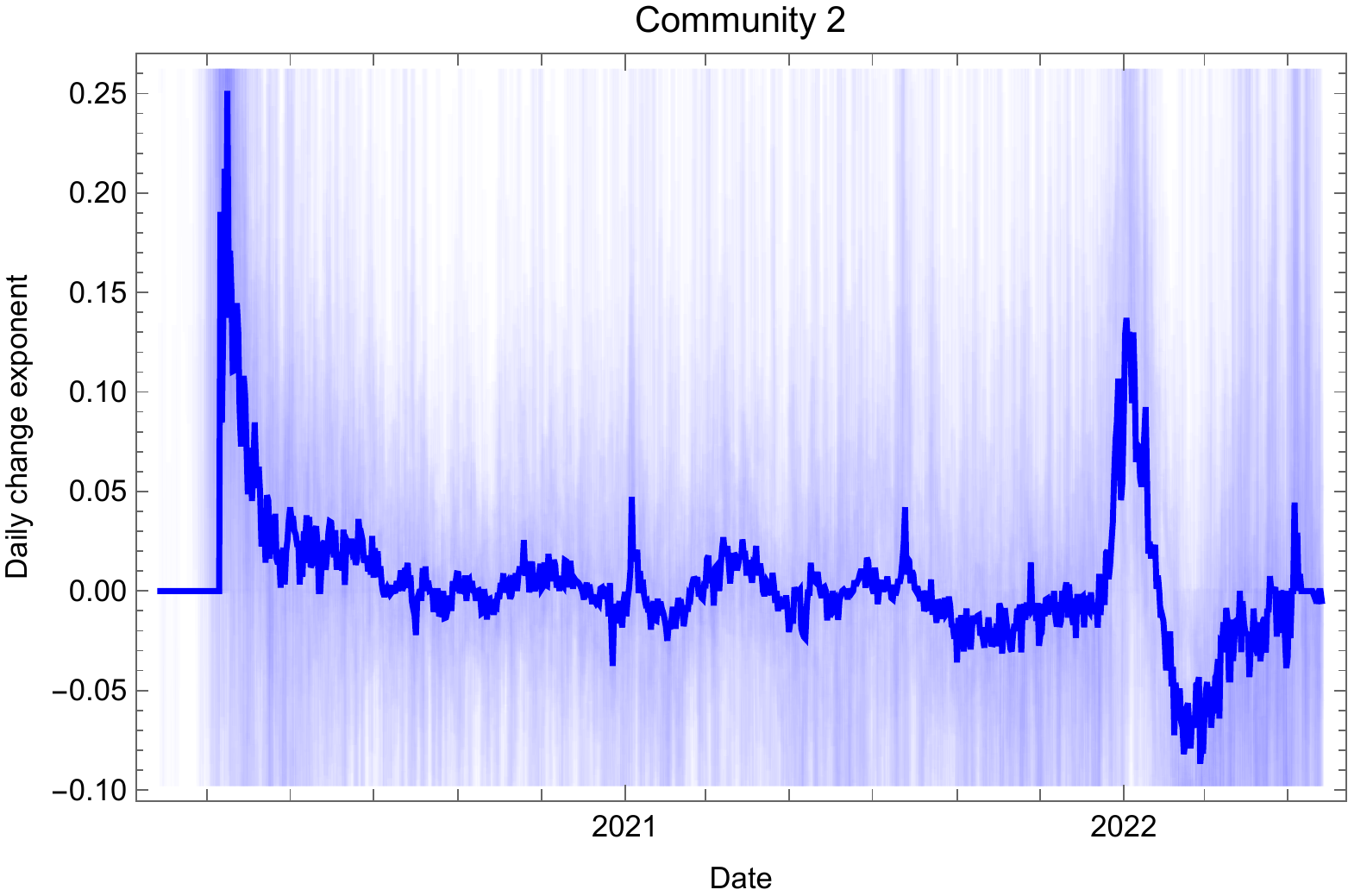} 
\vspace{0pt}
\end{minipage}
& ~ &
\begin{minipage}[t]{0.43\columnwidth}
\vspace{13pt}
{\scriptsize {\bf Community 2:} Afghanistan, Algeria, Armenia, Azerbaijan, Bahrain, Bangladesh, Bolivia, Bosnia and Herzegovina, Brazil, Chile, Colombia, Cuba, Dominican Republic, Ecuador, Ghana, Guatemala, India, Indonesia, Iraq, Israel, Japan, Kazakhstan, Kosovo, Kuwait, Mexico, Morocco, Nepal, Pakistan, Panama, Peru, Philippines, Russia, Saudi Arabia, South Africa, Taiwan, Turkey, Ukraine, United Arab Emirates, Uzbekistan, Venezuela, West Bank and Gaza}
\vspace{0pt}
\end{minipage}
\\
\begin{minipage}[t]{0.55\columnwidth}
\vspace{0pt}
\includegraphics[width=\columnwidth]{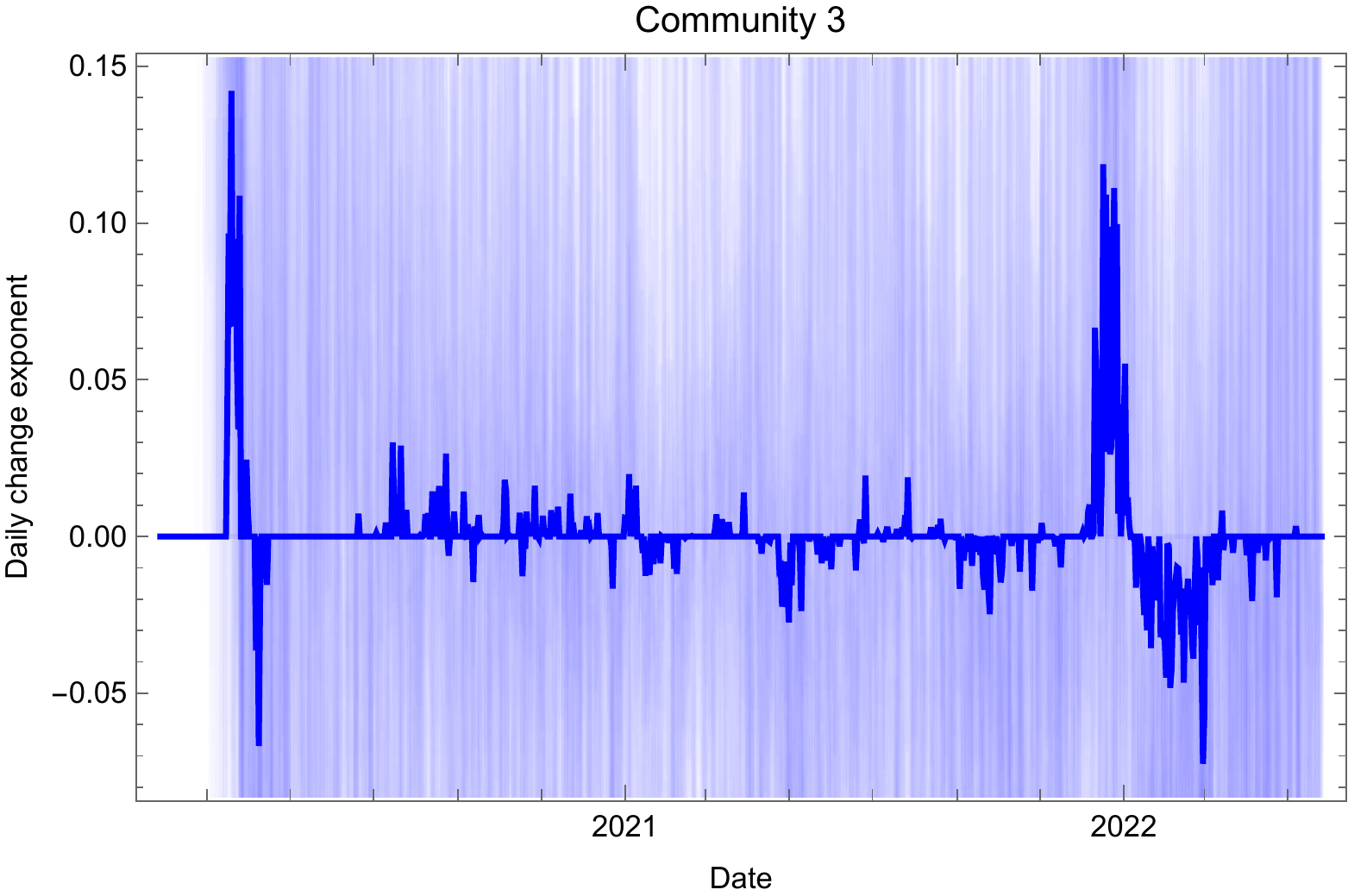} 
\vspace{0pt}
\end{minipage}
& ~ &
\begin{minipage}[t]{0.43\columnwidth}
\vspace{0pt}
{\scriptsize {\bf Community 3:} Argentina, Australia: Australian Capital Territory, Australia: New South Wales, Australia: Queensland, Australia: South Australia, Australia: Tasmania, Belarus, Botswana, Cameroon, Canada: Manitoba, Canada: Ontario, Canada: Quebec, Canada: Saskatchewan, Costa Rica, Cyprus, El Salvador, Ethiopia, France: Guadeloupe, France: Reunion, Honduras, Iceland, Jamaica, Jordan, Kenya, Laos, Libya, Maldives, Mauritius, Moldova, Montenegro, Mozambique, Namibia, Nigeria, Paraguay, Rwanda, Tunisia, Uganda, Uruguay, Zambia, Zimbabwe}
\vspace{0pt}
\end{minipage}
\end{tabular}
\caption{Time series plots of the daily change exponents for the three major communities. Thin blue lines show the exponent time series of individual countries (listed on the right), whereas thick blue lines show the median trend curve for each community. When the exponent was positive (negative) the number of daily new positive cases was growing (shrinking). Note that the actual peaks of the COVID-19 activities were where the median trend curve {\em switched its sign} from positive to negative, and {\em not} where the curve peaked in these plots.}
\label{timeseries}
\end{figure}

The modularity maximization method revealed three major communities in the global COVID-19 activity correlation network. Figure \ref{network-communities} visualizes the detected community structure in a network diagram, and Figure \ref{timeseries} shows time series plots of daily change exponents for the three major communities. The first community (Community 1) consists mostly of Europe, North America, and Southeast Asia, which showed similar six-peak patterns during the pandemic (Fig.~\ref{timeseries} top). The second community (Community 2) contains about half of the rest of the countries/regions, including Near/Middle East, Central and South Asia, and Central and South America, that loosely followed Community 1 but had a notable increase of activities because of the Delta variant in late 2020/early 2021 and was later impacted significantly by the Omicron variant in late 2021/early 2022 (Fig.~\ref{timeseries} middle). The third community (Community 3) was largely made of Africa, Central and East Canada, and the majority of Australia, which did not have major activities after the initial spike until another huge spike was caused by the Omicron variant in late 2021 (Fig.~\ref{timeseries} bottom). In this particular result, there were also two much smaller communities detected (Communities 4 and 5) but these smaller communities were not robust, as described below. 

\begin{SCfigure}[][tbp]
\includegraphics[height=\textheight]{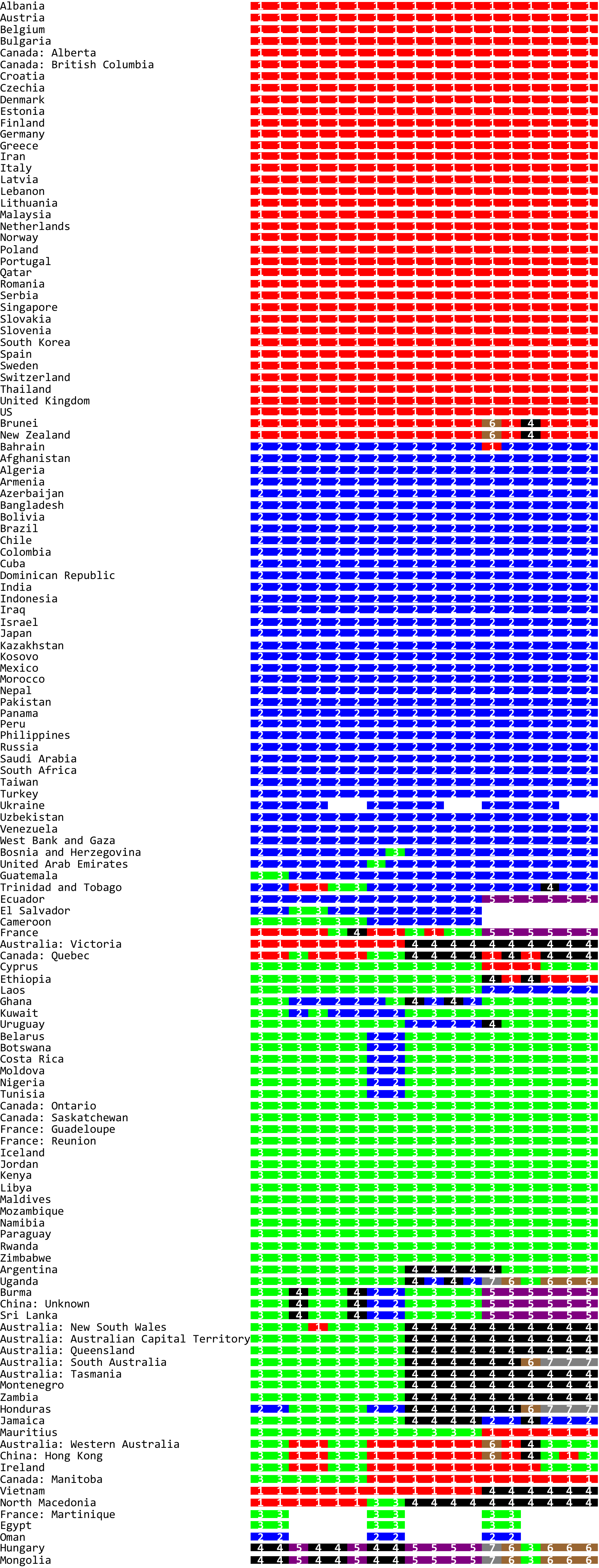}
\caption{Results of robustness testing experiments. Each row represents a country/region, whereas each column shows the result of network construction and community detection under a specific setting. A total of 18 setting variations were tested ($\rho: \{0, 0.05, 0.1\} \times \alpha: \{5, 7, 9\} \times \text{correlation}: \{\text{Pearson}, \text{cosine}\}$). The third column of the results corresponds to the main result described mostly in this paper ($\rho=0$, $\alpha=7$, correlation: Pearson). Members of Communities 1, 2, and 3 are shown in red, blue, and green, respectively. Members of other communities are shown in other colors. Blank space means that the country/region was not included in the correlation network. Countries/regions (rows) were ordered heuristically to maximize the visual consistency of community patterns.}
\label{setting-variations}
\end{SCfigure}

To test the robustness of observations, we also conducted the same community detection task with varied settings, including the correlation threshold for edge creation $\rho$ ($\rho \in \{0, 0.05, 0.1\}$), the daily change exponent's clipping range $\alpha$ ($\alpha \in \{5, 7, 9\}$), and the type of correlation measurements (Pearson correlation or cosine similarity). Figure \ref{setting-variations} shows the results. The three major communities were robustly detected under those varied settings, as seen in the figure as the three major color regions (red, blue, and green) that appeared consistently across settings (columns). Some countries/regions switched their community memberships (especially in Community 3; e.g., the regions in Australia), but the overall structure of those three major communities remained fairly consistent across the varied settings.

\begin{figure}[tp]
(a)\\
\hspace*{-20pt}\includegraphics[width=1.1\columnwidth]{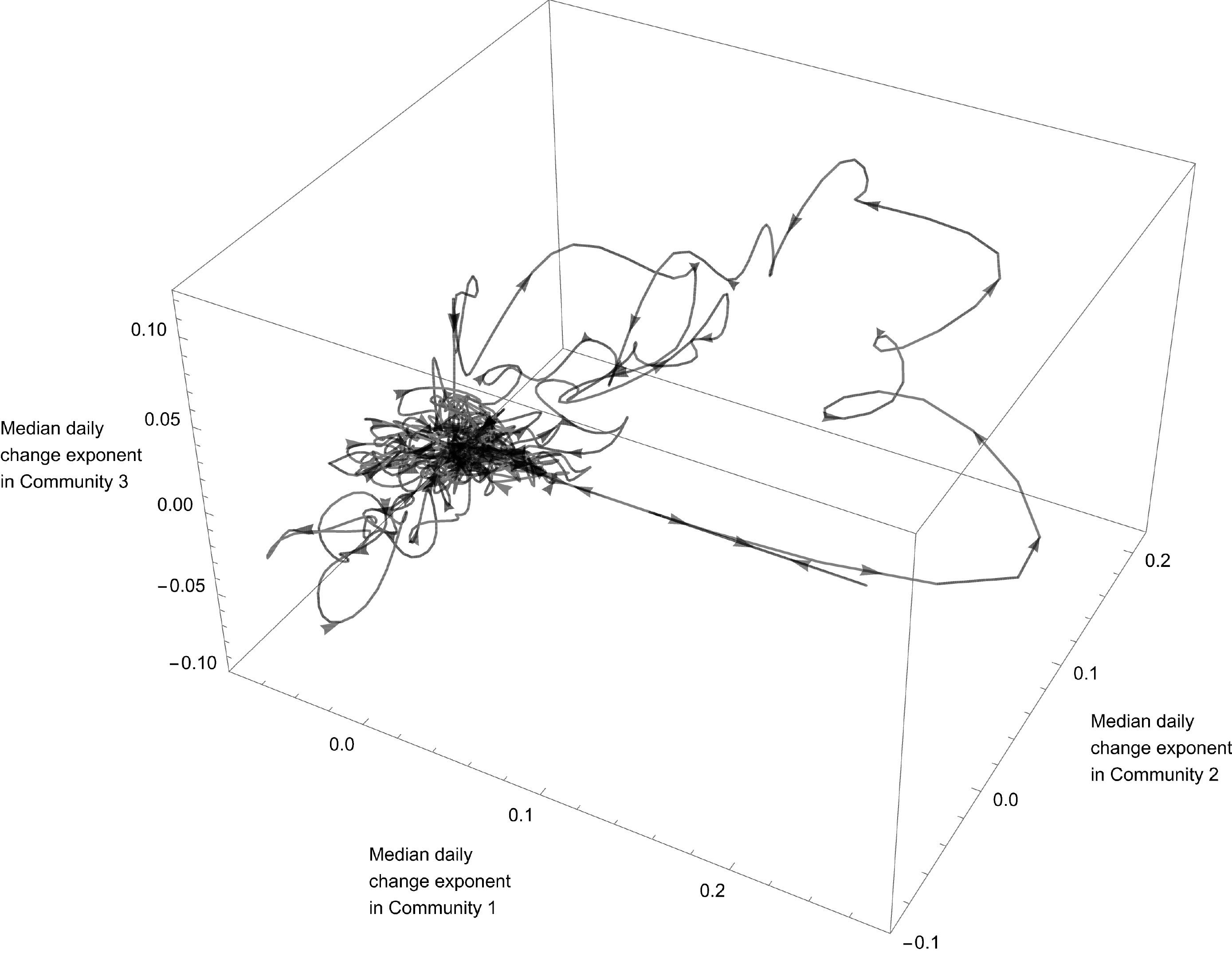}\\
(b) \hspace*{0.48\columnwidth} (c)\\
\includegraphics[width=0.48\columnwidth]{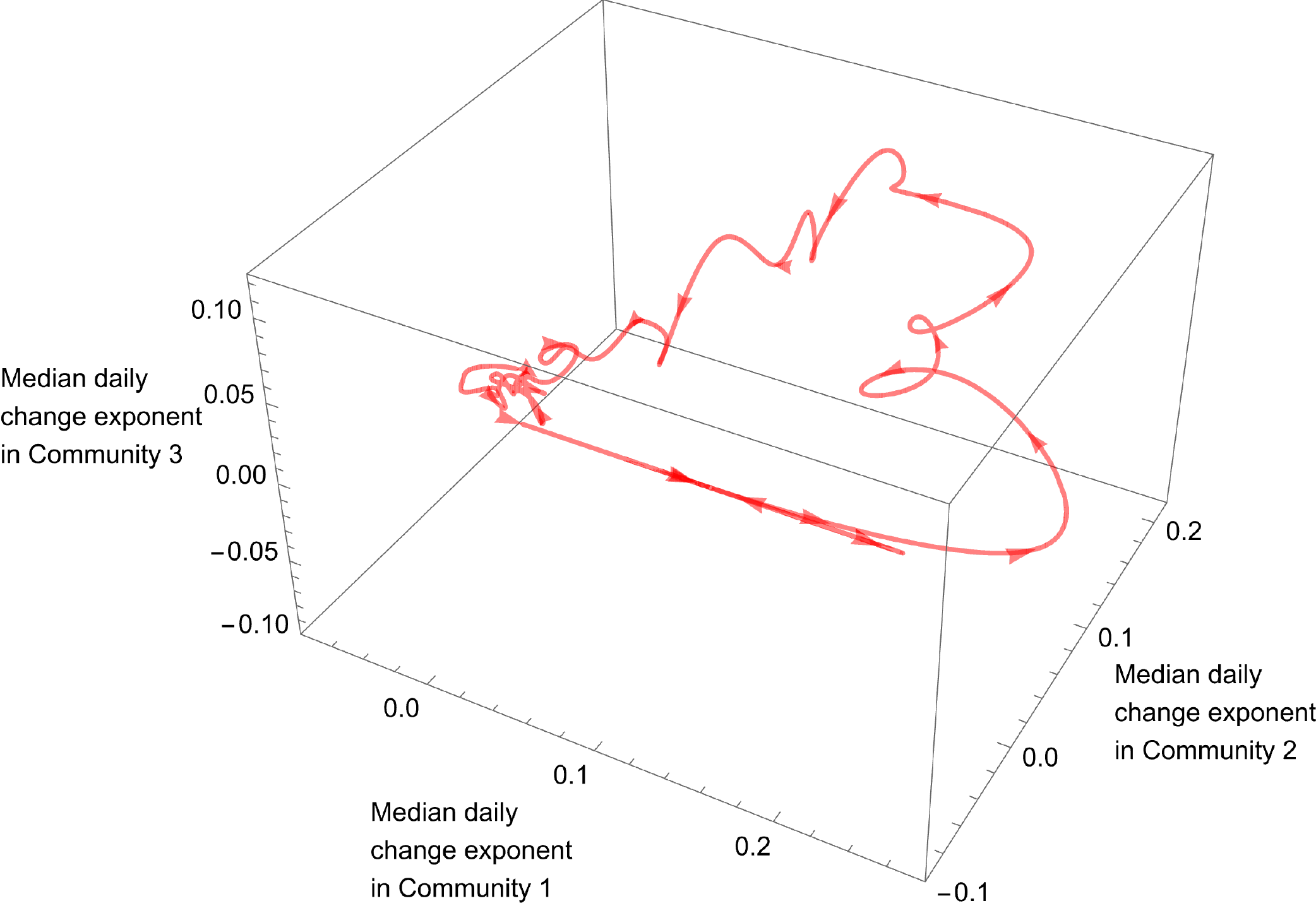}
\includegraphics[width=0.48\columnwidth]{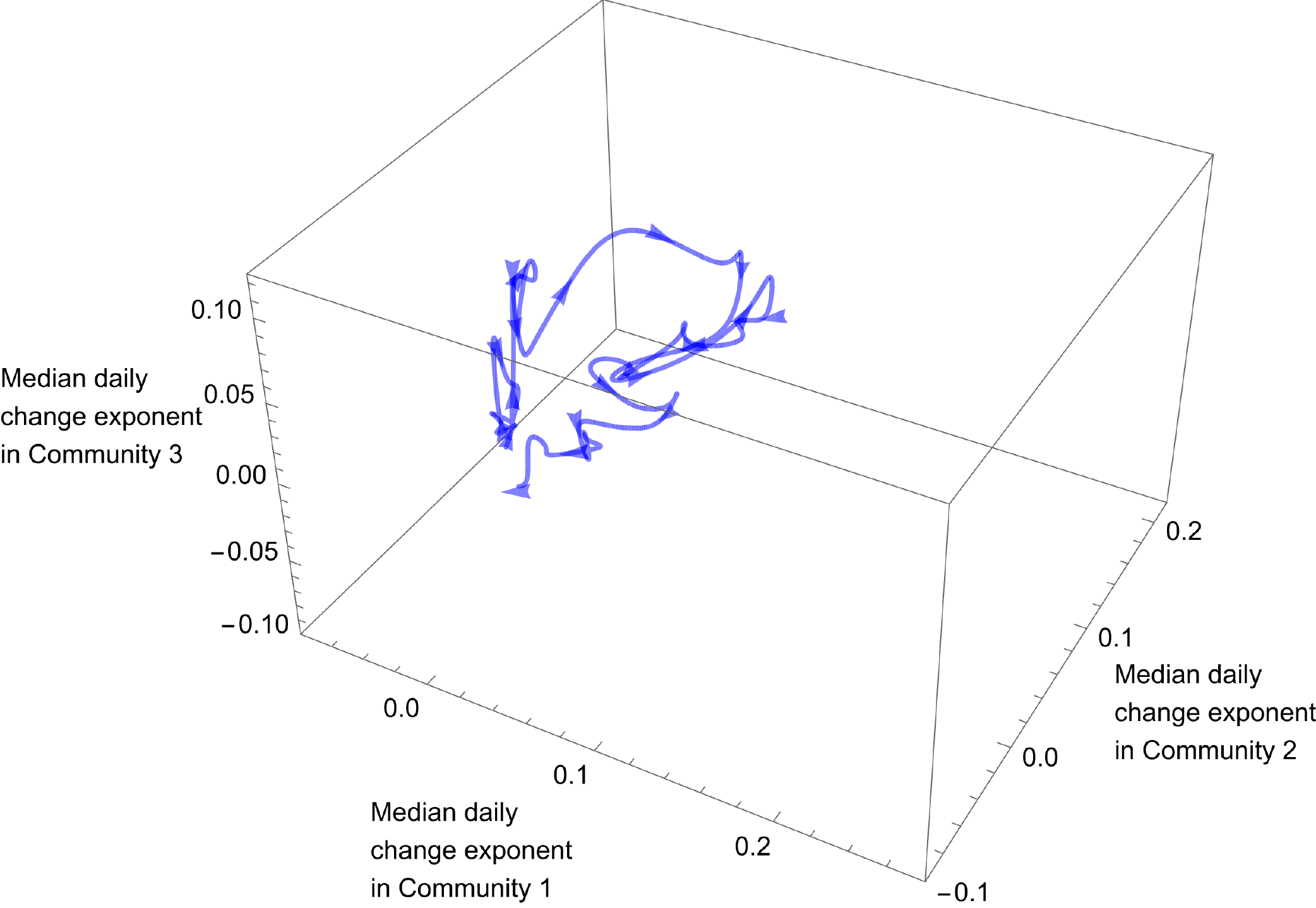}
\caption{Three-dimensional ``phase space'' visualization of the global pandemic state. The trajectory was constructed using the median trend curves of the daily change exponents for the three major communities (Communities 1, 2, and 3 as shown in Fig.~\ref{timeseries}) for $x$, $y$, and $z$ coordinates, respectively. The trajectory was smoothed using the B-spline algorithm for visualization purposes. Arrowheads indicate the direction of time. (a) Trajectory for the entire period of collected data. (b) Initial outbreak in January--April 2020, in which the activities spread in a sequence from Community 1 to Community 2, and then to Community 3. (c) Outbreak of the Omicron variant in December 2021 and January 2022, in which the activities spread in a different direction from Community 3 to Communities 1 and 2.}
\label{phase-space}
\end{figure}

The fact that the three major communities were robustly detected in the COVID-19 activity correlation network suggests that we may be able to use the aggregated activity levels within each of those communities as reasonable summary variables to capture the global pandemic dynamics. We therefore used the median curves of the daily change exponents for Communities 1, 2 and 3 (as shown in Fig.~\ref{timeseries}) for $x$-$y$-$z$ coordinates, respectively, to construct a three-dimensional ``phase space'' visualization of how the global COVID-19 pandemic progressed over time (Fig.~\ref{phase-space}). We applied the B-spline algorithm to smooth the trajectory to improve visual readability of the trajectory. Figure \ref{phase-space}(a) presents the overall trajectory of COVID-19 activities for the entire period of collected data, showing that the trajectory usually fluctuated around the origin ($(x, y, z) = (0, 0, 0)$) for the most of the time, but it also deviated significantly at several occasions when a major outbreak occurred. The largest loop that extends toward the right hand side (Fig.~\ref{phase-space}(b)) corresponds to the initial outbreak in early 2020, in which the COVID-19 activities spread in a sequence from Community 1 to Community 2, and then to Community 3, which is seen as a counter-clockwise rotation in Fig.~\ref{phase-space}(b). Meanwhile, the spread of the Omicron variant in late 2021/early 2022 is seen as a smaller extended loop at the center (Fig.~\ref{phase-space}(c)), in which the activities spread in a different direction from Community 3 to Communities 1 and 2. These visualizations offer an effective low-dimensional summary of complex global pandemic dynamics, which was made possible by detecting the global community structure in the COVID-19 activities.

\section{Conclusions}

In this study, we used the Johns Hopkins University's COVID-19 data to measure similarities of activity patterns between countries/regions over the last 2.5 years and constructed a global correlation network. We chose to characterize epidemic dynamics with the daily change exponents, instead of actual numbers of positive cases, so that the activity patterns could be captured and compared for a wide range of activity amplitudes that greatly varied from place to place and from time to time. Our network analysis robustly detected three major communities, each of which contained geographically distant countries/regions that exhibited similar COVID-19 activity history. This allowed for generating a trajectory of the global pandemic in a low-dimensional ``phase space'' by using the three median trend curves obtained from the three major communities as 3-D coordinates. This produced an effective summary and visualization of how the pandemic unfolded differently at different times, such as the initial outbreak and the more recent outbreak of the Omicron variant.

This study remains entirely empirical and observational, and it still leaves many open research questions. First, we have not gained any insight into what makes each of the three major communities a ``community,'' or in other words, whether there were sociopolitical, cultural and/or economic factors that were shared among the members of each community. Second, the network was constructed merely based on simultaneous similarity of the COVID-19 activity patterns, and it was not clear whether detecting more causal linkages with time delay (e.g., Granger causality, transfer entropy) would produce a different picture of the global community structure. Third, consideration of the possible temporal dynamics of network structures, such as the difference of global community structures between different seasons/years, were completely omitted. Fourth, we studied only the dataset of the recent COVID-19 pandemic, and thus it was unclear whether the detected global community structure would apply only to this particular pandemic or to other global infectious diseases (e.g., influenza) more generally. Finally, we have not yet explored how this global network modeling and analysis could inform epidemic prediction and mitigation efforts at local/regional levels. We hope this study inspires further future research on these and other questions about the global network of epidemic dynamics.

\section*{Acknowledgments}

We thank three anonymous reviewers for their constructive comments that were very helpful in improving the clarity of this manuscript.

\end{document}